\documentstyle[prb,aps]{revtex}
\twocolumn
\begin{document}
\input{psfig}
\small
\draft
\title
{\bf Single and double bit quantum gates by manipulating degeneracy}
\author{T. Hakioglu$^{(1)}$
, J. Anderson$^{(2)}$
, F. Wellstood$^{(2)}$}

\address{1. Department of Physics, Bilkent University, Bilkent, 06533
Ankara, Turkey
\\
2. Department of Physics, University of Maryland, College Park,
MD 20742}

\maketitle
\twocolumn
\begin{abstract}

A novel mechanism is proposed for single and double qubit state
manipulations in quantum computation with four-fold degenerate
energy levels. The principle is based on starting with a four-fold
degeneracy, lifting it stepwise adiabatically by a set of control
parameters and performing the quantum gate operations on
non-degenerate states. A particular realization of the proposed
mechanism is suggested by using inductively coupled rf-squid loops
in the recently observed macroscopic quantum tunneling regime where the
energy eigen levels are directly connected with the measurable
flux states. The one qubit and two qubit controlled operations are
demonstrated explicitly. The appearance of the flux states 
also allows precise read-in and read-out operations by the  
measurement of the flux.
\end{abstract}

\pacs{PACS numbers:}

\narrowtext

Recent advances on the experimental efforts to demonstrate
the fundamental single and two bit quantum gates by ion
trap\cite{iontrap} and nuclear magnetic resonance (NMR)
experiments\cite{nmr} as well as the observation of the 
coherent Rabi oscillations in Cooper pair boxes\cite{cpb} are hinting at
new challenges awaiting the realization of the quantum
computational devices. It has been shown theoretically that the basic 
gate operations based on the unitary transformations of the qubits can
be performed by single and double qubit
manipulations\cite{universal}. Therefore it is a necessary first
step that a proposed mechanism should demonstrate these
fundamental operations as well as a successful read-in and
read-out before any decoherence comes into play. A  
next step which is not detailed in this letter is to construct
arrays of independent qubits or coupled qubit pairs to perform
parallel gate operations. In our suggested mechanism the state space is 
four dimensional corresponding to the state space of a pair of qubits. 
The paired
qubit states are inseparable; nevertheless, each qubit can be
manipulated individually. The mechanism also allows controlled
operations on selected bits rather easily without affecting those
bits that need to be unchanged. A more detailed discussion of the
proposed mechanism with its specific experimental realizations
using rf-squid loops will be published elsewhere.

\subsection{The mechanism}
The mechanism is  based upon coupling a pair of ideal two level
systems with manifestly degenerate energy levels at the energy $E_0$ by two
complex coupling parameters as indicated in Fig.1a. The coupling
${\cal C}_1$ lifts the four-fold degeneracy to two-fold
(Fig.\,1b). The second coupling ${\cal C}_2$ couples the two
remaining degenerate levels in each pair and lifts the degeneracy
completely (Fig.1c). The states in the initial four-fold
degenerate configuration at ${\cal C}_1={\cal C}_2=0$ are labelled
by $\vert m\, n\rangle,~(m,n=0,1)$. We consider them to be the
eigen states of some underlying non interacting system. In the
qubit language below the first (second) bit is $m \, (n)$. The
states $\vert m\,n\rangle$ are the components of the four 
dimensional row vector $(\vert
11\rangle,\vert 00\rangle, \vert 10\rangle, \vert 01\rangle)$ which 
is to be considered as the basis for the model Hamiltonian. The Hamiltonian
including the degeneracy breaking couplings is given in this four 
dimensional basis by the Hermitian matrix  
\begin{equation}
{\bf H}_{{\cal C}_1,{\cal C}_2}=\pmatrix{E_0 &
         {\cal C}_1 & {\cal C}_2 & 0\cr
         {\cal C}_1 & E_0  & 0 & {\cal C}_2 \cr
         {\cal C}_2 & 0 & E_0 & {\cal C}_1\cr
         0 & {\cal C}_2 & {\cal C}_1 & E_0\cr}~. 
\label{prl.1}
\end{equation}
Without loss of generality we assumed that the couplings are 
{\it real and positive}  
and that they are applied adiabatically in the order ${\cal C}_1$
first and ${\cal C}_2$ second (the order is commutative) with 
$0 < {\cal C}_2 \le {\cal C}_1$. The  
diagonalizing matrices for independent couplings are unitary
transformations acting on the four dimensional basis. These are 
${\cal U}_1$ for ${\cal C}_1 \ne 0, {\cal C}_2=0$ and ${\cal U}_2$ 
for ${\cal C}_1=0, {\cal C}_2\ne 0$ as given by 
\begin{equation}
{\cal U}_1={1 \over \sqrt{2}}\,\pmatrix{1 & 1 & 0 &0\cr
                    1 & -1 & 0 & 0\cr
                    0 & 0 & 1 & 1 \cr
                    0 & 0 & 1 & -1\cr}~,\qquad
{\cal U}_2={1 \over \sqrt{2}}\,\pmatrix{1 & 0 & 1 &0\cr
                    0 & 1 & 0 & 1\cr
                    1 & 0 & -1 & 0 \cr
                    0 & 1 & 0 & -1\cr}~. 
\label{prl1b}
\end{equation}
When both couplings are turned on, the Hamiltonian (\ref{prl.1})
is diagonalized by ${\cal U}={\cal U}_2 \, {\cal U}_1$ as ${\bf
H}_d=diag\{ E_0+ {\cal C}_1+{\cal C}_2,
E_0+{\cal C}_1-{\cal C}_2, E_0-{\cal
C}_1+{\cal C}_2, E_0-{\cal C}_1
-{\cal C}_2\}$ which correspond to the levels
in Fig.1c. For arbitrary couplings ${\cal C}_1,{\cal C}_2 \ne 0$
the eigen energies are all nondegenerate and they are distributed
symmetrically around the major center $E_0$ and the two minor ones
$E_0 \pm {\cal C}_1$. The single and double qubit
quantum gate operations are performed ideally by manipulating the
strengths of at least one of the coupling constants
and making use of the invariance of the major center under
adiabatic changes of both couplings and of the two minor centers under
that of ${\cal C}_2$.
In Fig.1c we have indicated the eigenstates with up and down arrows. 
In the double-SQUID model considered below they correspond to the 
flux states. These states are  
connected with the initial basis states $\vert m\,n\rangle$ by
\begin{equation}
\pmatrix{\vert \uparrow \uparrow \rangle \cr \vert \downarrow \downarrow
\rangle \cr
   \vert \uparrow\downarrow\rangle \cr \vert \downarrow\uparrow
\rangle\cr}={\cal U}_2\,{\cal U}_1\,\pmatrix{\vert 11\rangle \cr 
                                             \vert 01\rangle \cr
                                             \vert 00\rangle \cr
                                             \vert 10\rangle\cr}~. 
\label{updownst.1}
\end{equation}
Note that the ordering of the $\vert m\,n\rangle$ basis  
is made with respect to the ordering of the energy eigen values of the 
flux states on the left hand side in (\ref{updownst.1}) which is  
different than  
the ordering used for constructing the Hamiltonian matrix (\ref{prl.1}). 
For later convenience we introduce the short notation
\begin{equation}
\vert {\bf V}\rangle=
\pmatrix{\vert \uparrow \uparrow \rangle \cr 
         \vert \downarrow \downarrow \rangle \cr 
         \vert \uparrow\downarrow\rangle \cr 
         \vert \downarrow\uparrow \rangle\cr}~,\qquad 
\vert {\bf v}\rangle =
\pmatrix{\vert 11\rangle \cr
         \vert 01\rangle \cr
         \vert 00\rangle \cr
         \vert 10\rangle\cr}
\label{notation}
\end{equation}

\subsection{An rf-SQUID realization of the mechanism}
Physically, this suggested four state mechanism can be realized by
using inductively coupled rf-squids as depicted in Fig. 2. The
SQUIDs must operate in the macroscopic quantum tunneling (MQT)
regime\cite{MQT,Cat} which requires temperatures on the order of mili
Kelvin  for standard Josephson junctions.
The four level system of Fig.1a is constructed by preparing two
identical SQUIDs A and B in which each qubit is represented, to a
good approximation, by its two lowest energy eigen states (denoted 
as $\vert 0\rangle$ and $\vert 1\rangle$); in the MQT regime these
are nearly degenerate (pseudo-degenerate) in energy (due to the 
fluctuations in the flux) compared to the next (third) level.  

A qubit which is close to being ideal  
is characterized by an isolated hypothetical 
two-level system with a high qubit quality ratio 
$\eta=(E_2-E_1)/(E_1-E_0)$, where $E_0$ and $E_1$ are the two
lowest levels and $E_2$ is the next (well-separated) level. In the
case of manifest degeneracy, i.e. $E_0=E_1 \ne E_2$ one obtains
$\eta=\infty$, the ideal case. For the realistic case here using coupled 
SQUIDs, we search for a pseudo degenerate configuration (by performing a 
numerical search at different parameter values) in which 
we obtain sufficiently high qubit   quality factors on the order
of $\eta \simeq 40$. In Fig.3 the calculated wavefunctions for the
two lowest states of a symmetric double-well single-SQUID potential is plotted
for the specific parameter values which are junction capacitance
 $C=0.4 pF$, 
the loop inductance $L=100 pH$ and the potential modulation
parameter\cite{Cat} $\beta_L=2\pi I_c I/\Phi_0=1.1$ where $I_c$
is the critical Josephson current, $I$ is the superconducting
current in the loop and $\Phi_0$ is the superconducting flux
quantum $\Phi_0=h c/2e$. 
In the numerical calculations the range of the  
junction and the loop parameters are chosen such that the pseudo-degenerate   
eigen states are located in the double well, whereas the third level is above  
the central maximum of the double well potential. 
For the flux dominated regime the Josephson energy $E_J$ is expected to be  
much larger than the charging energy $E_c$ in the junctions. 
Considering that the sum of these energies is fixed   
($E_T=E_c+E_J$ is the total energy in the noninteracting SQUID loop), we find 
for our parameters $E_J/E_T \sim 95 \%$ for the symmetric configuration 
of the double well potential. If the symmetric configuration is tilted, 
the additional level splitting raises the charging energy. For the range of 
interest of the bias flux (see Fig. 4) a drop to $90 \%$ in the above ratio 
is observed as a result of the tilting.   
The pseudo- degenerate single SQUID states $\vert 0\rangle$
 and $\vert 1\rangle$ have respectively even and odd
parities. We will refer to these states as the one-bit parity
basis. The odd and even superpositions of the parity states are
given by $\vert \uparrow \rangle= (\vert 0\rangle+\vert
1\rangle)/\sqrt{2}$ and $\vert \downarrow \rangle= (\vert
1\rangle-\vert 0\rangle)/\sqrt{2}$. These are the single-well
localized flux states. Hence, we refer to them as the one-bit flux
basis. The single-well localized means that these one-bit states
actually refer to a single SQUID where they are well confined to a
single well of the one-squid double-well potential. The calculated  
flux states corresponding to the chosen SQUID parameters above are
also shown in Fig.3 (the solid triangles).

We now change from the single bit-single SQUID to the double bit-
double SQUID configuration. The case for two identical SQUID loops in the
pseudo-degenerate regime with no mutual interaction is given by
Fig.1a. Inductively coupling these two identical SQUID loops in
Fig.2 by a weak inductance $M (\ll L)$, lifts the four-fold
degeneracy to two-fold degeneracy. In the coupled SQUID model the
weak mutual inductance cannot be varied after fabrication. This
mutual inductance corresponds to the coupling ${\cal C}_1$
(Fig.\,1) and we believe that $M \simeq 10^{-2}\,L$ is a
reasonable range to consider in the experiment. Once the SQUIDs
are manifactured, $M$ is fixed, therefore the four fold degenerate
configuration in Fig.1a is out of reach. However we still have
leverage on the second coupling. This second  
coupling is the tilt of the symmetrical Josephson potential
of the two SQUIDS simultaneously deriving both potentials to  
unsymmetric configurations by means of an additionally applied flux 
$\Delta \Phi_{ex}$ (The total flux is $\Phi_{ex}=\Delta \Phi_{ex}+\Phi_0$).  
This is modelled by the second coupling ${\cal C}_2$ which lifts the
remaining two-fold degeneracy completely. The combined effects of
the mutual inductance, which we now consider to be
$M=0.03\,L$, and the tilt can be calculated numerically by
diagonalizing the inductively coupled two-SQUID Hamiltonian. The
calculated eigenenergies are plotted against the tilting parameter
$y_{ex}=\pi\,\Phi_{ex}/\Phi_0$ in Fig.\,4. 
The configuration
$y_{ex}=\pi$ (i.e. $\Phi_{ex}=\Phi_0$)
  corresponds to the symmetric SQUID potential. The
solution of the Schr\"{o}dinger equation for this potential and  
this coupled SQUID configuration has two closely spaced energy
levels (at the mentioned parameter range above) comprising the
best pseudo-degenerate approximation we ever found to the manifest
double degeneracy in Fig.1b. At this point we read from the vertical axis 
in Fig.\,4 that ${\cal C}_1=10^{-4}\,eV$. As the double-well potentials are
tilted in parallel and simultaneously in both SQUIDs the shift
obtained for each eigenenergy is approximately linear in the bias
flux $\Phi_{ex}$. This indicates that the coupling can be primarily
represented as a first order perturbation in $\Phi_{ex}$. 
A comparison of the Fig.\,1 
with the calculations shown in Fig.\,4
indicates that the coupled rf-SQUID model is in essence an isolated 
four-level quantum system. 
Recently, Schr\"{o}dinger Cat states have been obtained experimentally 
by manipulating these two couplings in the MQT regime.\cite{Cat}
 For the physically relevant range $0 \le y_{ex} \le
y_{ex}^*$ the centers of the splitting energies in Fig.\,4 also respect 
the manifest conservation principle of the energy centers in Fig.\,1. 
In fact,  in the numerical calculations we found a shift   
smaller than $2\%$ in the energy centers for the whole range 
$0 \le y_{ex} \le y_{ex}^*$. 

In this paper we will refer to two different representations of
the basis states for two coupled SQUIDs. The first, which is the even
and the odd states of the mutually noninteracting SQUIDs, is
represented by the vector $\vert {\bf v} \rangle = \{\vert
11\rangle, \vert 00\rangle , \vert 10 \rangle,  \vert 01
\rangle\}$ as in section A. Here in the state $\vert m n \rangle$
$m$ represents the parity state of the SQUID A and $n$ represents that
of SQUID B. We will designate $\vert {\bf v}\rangle$ as the {\it two-bit
parity basis}. The second, which is the set of flux states for the
coupled and tilted SQUIDs , i.e. both couplings are nonzero, is
represented by $\vert {\bf V} \rangle = \{\vert \uparrow
\uparrow\rangle, \vert \downarrow\downarrow \rangle, \vert
\uparrow \downarrow\rangle, \vert \downarrow \uparrow\rangle \}$
which we refer to as the {\it two-bit flux basis}.

\subsection{The quantum gate operations}
According to the Hamiltonian (\ref{prl.1}), the eigen levels
 in Fig.1b corresponding to ${\cal C}_1 \ne 0$ and
 ${\cal C}_2=0$ are the doubly degenerate states $\vert {\bf v}^\prime\rangle
={\cal U}_1\,\vert {\bf v}\rangle$. Explicitly these are, $(\vert
00\rangle+\vert 11\rangle)/\sqrt{2}$ and $(\vert 10\rangle+\vert
01\rangle)/\sqrt{2}$ for $E_0+\vert{\cal C}_1\vert$, and, $(\vert
10\rangle-\vert 01\rangle)/\sqrt{2}$ and $(\vert 00\rangle-\vert
11\rangle)/\sqrt{2}$ for $E_0-\vert{\cal C}_1\vert$. If ${\cal
C}_2$ is also turned on as in Fig.1c, the non-degenerate energy
eigenstates become the two-bit flux states defined in 
Eq.\,(\ref{updownst.1}) and written explicitly as  
\begin{equation}
\pmatrix{\vert \uparrow \uparrow \rangle \cr \vert \downarrow \downarrow
\rangle \cr
   \vert \uparrow\downarrow\rangle \cr \vert \downarrow\uparrow
\rangle\cr}
={1 \over 2}\,\pmatrix{1 & 1 & 1 & 1\cr
                       1 & -1 & 1 & -1 \cr
                     1 & 1 & -1 & -1 \cr
                    1 & -1 & -1 & 1\cr}\,
\pmatrix{\vert 11\rangle \cr \vert 01\rangle \cr 
         \vert 00 \rangle \cr \vert 10\rangle\cr}~. 
\label{flux.1}
\end{equation}
The choice of the basis in which the fundamental quantum gates are
to be operated should be determined by the accessibility of the
input and output states by the measurement mechanism. This
condition actually means that all information in the relative
magnitudes and the phases of these states should be measurable.
The absolute values of the amplitudes of the flux states are
directly measurable whereas their relative phase factors do not
couple to the measurement (see below). 
Since it is crucial to be able to read and write the
relative phase between the flux states, we choose the two-bit
parity basis $\vert {\bf v} \rangle$ instead of the two-bit flux basis 
$\vert {\bf V} \rangle$ to perform   
the gate operations. The flux basis is used for the I/O operations. 
Before and after performing the logic gates it is therefore needed 
to switch between the $\vert {\bf v}\rangle$ and the 
$\vert {\bf V}\rangle$ bases. 
Notice that the matrices ${\cal U}_1$ and ${\cal U}_2$ are two-bit 
Hadamard transformations. In particular, it can be verified that  
\begin{equation}
{\cal U}_1={\bf 1} \otimes \fbox{H}~,\qquad {\cal U}_2=\fbox{H} \otimes 
{\bf 1}~,\quad \fbox{H}={1 \over \sqrt{2}}\pmatrix{1 & 1\cr 1 & -1\cr}
\label{2bithad}
\end{equation}
where ${\bf 1}$ is the $2\times 2$ unit matrix. The matrix $\fbox{H}$ is 
the one bit Hadamard transformation. The transformation from the natural 
eigenbasis $\vert {\bf V}\rangle$ to the parity basis 
$\vert {\bf v}\rangle$ can therefore be performed by two successive 
two-bit Hadamard transformations. 
In this way any phase information of a quantum state encoded in
$\vert {\bf v}\rangle$ basis contributes to the relative amplitudes
of the $\vert {\bf V}\rangle$ basis yielding direct accesibility to the
measurement. This transformation can be summarized as 
$\vert {\bf V}\rangle ~\stackrel{{\cal U}_1}{\longrightarrow}~ 
\vert {\bf v}^\prime \rangle 
 ~\stackrel{{\cal U}_2}{\longrightarrow}~ 
\vert {\bf v} \rangle$. 
The effective result of this
transformation is that each state in the parity basis is finaly
assigned a distinct eigen energy as shown below. 
This transformation is equivalent to effectively assigning each component 
$\vert m\,n\rangle$ of the $\vert {\bf v}\rangle$ basis an energy eigenvalue. 
The logic operations are then performed on the $\vert {\bf v}\rangle$ basis. 
We now demonstrate the fundamental
single-qubit operations, i.e. the phase flip  and
the Hadamard transformation.

{\it Single qubit operations:}

Although the mechanism is based on four states, it is possible to
identify two individual flux qubits and perform single bit
operations on each one independently. 
In the parity basis, $\vert
m\,n\rangle$ we assign the first bit as the control (qubit no.1) 
and the second
bit as the target (qubit no.2). By definition, the single qubit 
operations are performed on 
the target in the four dimensional state space unconditionally
from the state of the control bit. Whereas the controlled operations
(two-qubit operations) are performed when the control bit is in a
particular, i.e ($0$ or $1$) state. With this convention we now
start with the single qubit operations.

{\bf I.} The phase flip is a special case of the phase evolution
described by
\begin{eqnarray}
\fbox{Z}_{\phi}: \pmatrix{\vert 11\rangle \cr \vert 01\rangle \cr
                   \vert 00\rangle \cr \vert 10\rangle\cr}~\mapsto
\pmatrix{&\vert 11\rangle & \cr &\vert 01\rangle & \cr
                   e^{-i2\phi}& \vert 00\rangle & \cr
                   e^{-i2\phi}&\vert 10\rangle &\cr}
\label{z.1a}
\end{eqnarray}
 The phase flip is obtained in (\ref{z.1a})
at $\phi=\pi/2$ which implies that the state of the second bit is changed as 
$1 \to 1$ and $0 \to -0$ independent from the state of the first bit. 
It is shown below that this can be obtained as a direct sum of two controlled
phase flips.

The phase evolution $\fbox{Z}_{\phi}$ in Eq.\,(\ref{z.1a}) is achieved  
in the two coupled SQUID realization in two steps. We switch the coupling 
configuration of the system to ${\cal C}_2={\cal C}_1$ 
which makes the states $\vert 01\rangle$ and $\vert
00\rangle$ degenerate at the major center $E_0$.   
By this switching, the remaining (nondegenerate) states, i.e. $\vert 11
\rangle$ and $\vert 10 \rangle$ are now at the energies 
$E_0 +2 {\cal C}_1$ and  
$E_0 -2 {\cal C}_1$ respectively.  The four state system is then permitted 
to freely evolve in this configuration. Degenerate 
states do not acquire any relative phase with respect to each other. 
At this first stage the phases acquired by the states are described by
\begin{eqnarray}
\pmatrix{\vert 11\rangle \cr \vert 01\rangle \cr
                   \vert 00\rangle \cr \vert 10\rangle\cr}~\mapsto
e^{iE_0\,t}\,
\pmatrix{e^{i\phi}&\,\vert 11\rangle & \cr &\,\vert 01\rangle & \cr
                   &\,\vert 00\rangle & \cr
                   e^{-i\phi}&\vert 10\rangle &\cr}
\label{z.1b}
\end{eqnarray}
where $\phi=2 {\cal C}_1\,t$ as the time dependent
phase. In the second stage we set ${\cal C}_2=-{\cal C}_1$ which
turns the states $\vert 11\rangle$ and $\vert 10\rangle$ to become 
degenerate at the energy $E_0$ and $\vert 01\rangle$ and $\vert
00\rangle$ nondegenerate at energies $E_0 +2 {\cal C}_1$
and $E_0 -2 {\cal C}_1$ respectively. An identical phase
evolution is then performed on the states $\vert 01\rangle$ and
$\vert 00\rangle$. The net transformation is given by
\begin{eqnarray}
\pmatrix{\vert 11\rangle \cr \vert 01\rangle \cr
                   \vert 00\rangle \cr \vert 10\rangle\cr}~\mapsto
e^{i2\,E_0\,t}\,
\pmatrix{e^{i\phi}&\,\vert 11\rangle & \cr e^{i\phi}&\,\vert 01\rangle & \cr
                   e^{-i\phi}&\,\vert 00\rangle & \cr
                   e^{-i\phi}&\vert 10\rangle &\cr}
\label{z.1c}
\end{eqnarray}
which, after factoring out the overall phase $e^{i2(E_0 \,t+\phi)}$ 
is equivalent to Eq.\,(\ref{z.1a}). The phase flip
is obtained at $\phi=\pi/2$ corresponding to $t_1=\pi/4{\cal C}_1$
which can be described as
\begin{eqnarray}
\fbox{Z}_{\pi/2}=&~& \Bigl[-i\, e^{-i\,E_0\,t_1}\,e^{i{\bf H}_{{C}_1,-{\cal
C}_1}\,t_1}\Bigr]\, ({\cal C}_2 \to -{\cal C}_1) \nonumber \\
&\times &\Bigl[-i\,e^{-i\,E_0\,t_1}\,e^{i{\bf H}_{{\cal
C}_1,{\cal C}_1}\,t_1} \Bigr]
({\cal C}_1 \to {\cal C}_1)~. 
\label{z.3}
\end{eqnarray}
where the action of the operators (matrices) is defined rightwards. 
Notice that, Eq.\,(\ref{z.3}) is manifestly the composition of two
controlled phase flips where these two operations commute.
Subtracting the overall phase, the first square bracket is equivalent to
[$\vert 11\rangle ~\to ~\vert 11 \rangle, \vert 10\rangle ~\to
~-\vert 10\rangle$]. This is the controlled phase flip $1 \to 1$
and $0 \to -0$ when the control bit is in the state $1$. Similarly,
the second bracket denotes the same controlled phase flip 
when the control bit is in
the $0$ state. Hence, this method proves an efficient realization
of all controlled operations as well (see the double qubit operations
below).

{\bf II.} The Hadamard (\fbox{H}) transformation is a special case that can
be obtained by rotations and phase flips on the Bloch sphere and
its realization is very similar to the one discussed already in
the phase flip. The idea is to perform two consequtive 
controlled rotations by coupling the non-degenerate states to an  
external pulse resonant at the energy difference $4{\cal C}_1$. 
Basically the two-bit Hadamard gate can be obtained in the following steps
\begin{eqnarray}
&&\to ({\cal C}_2 ~\to ~{\cal C}_1)\, \nonumber \\
&&~~~\to [R_z(\pi/4)\,R_x(\pi/4)\,R_z(\pi/4)] \nonumber \\
&&~~~~~~\to  [{\cal C}_2 ~\to ~ -{\cal C}_1]  \nonumber \\
&&~~~~~~~~~\to  [R_z(\pi/4)\,R_x(\pi/4)\,R_z(\pi/4)] 
\label{had.1}
\end{eqnarray}
In the first step $[{\cal C}_2 ~\to ~{\cal C}_1]$ 
the inner states, i.e. $\vert 01\rangle, \vert
00\rangle$, are made degenerate at the major center $E_0$. 
In the second step
a Hadamard transformation is performed (up to an overall phase
$e^{-i\pi/2}$) on the outer states $\vert 11\rangle$ and $\vert 10\rangle$ 
(for those the first bit is $1$) when the inner states (the first bit is zero) 
are degernerate. 
In this term, the states $\vert 11\rangle$ and $\vert 10\rangle$
are rotated by $R_x(\pi/4)$ sandwiched
 between two phase evolutions at $\phi=2 \times \pi/4$
indicated by $R_z(\pi/4)$. The overall phase accumulated on the
four states is $e^{iE_0\,t_2}$, where $t_2$ is the duration of the
pulse given by $t_2=\pi/(4\kappa)$ with $\kappa$ describing the
coupling strength of the levels to the external pulse [The
coupling mechanism to the external field is similar to the ion
trap experiments and will not be 
repeated here]. The third factor $[{\cal C}_2 ~\to ~-{\cal C}_1]$
indicates that ${\cal C}_2$ is now switched in the opposite
direction swapping the inner and the
outer states. The fourth factor in (\ref{had.1}) is
another Hadamard transformation in the interchanged configuration 
(when the first bit is $1$).
The same front factor $e^{iE_0\,t_2}$ as in the step 1 also appears here 
as an overall phase. After the fourth step the inner and outer states
should be swapped back to the initial configuration. 
Note that in all single and
double qubit operations there is an overall phase acquired by all
qubit states. Also note that the \fbox{H} gate described by
Eq.\,(\ref{had.1}) is a product of the two transformations [indicated
by the square brackets therein] where each individual bracket  
is a controlled $\fbox{H}$ conditioned by the state of the first  
bit.

{\it Double qubit operations}:
The controlled phase flip was considered previously. It basically 
comprises the first or the second half of the $\fbox{Z}_{\pi/2}$ 
operation in Eq.\,(\ref{z.3}). We will therefore not consider it here again. 
The other crucial double qubit operation is the Controlled-NOT gate.

{\it Controlled NOT}

Ideally, the \fbox{CNOT} gate can be obtained from \fbox{CZ} and
\fbox{H} by\cite{universal,NChuang}
\begin{equation}
\fbox{CNOT}=\fbox{H}\,\fbox{CZ}\,\fbox{H}
\label{cz.3}
\end{equation}
It is much easier to obtain \fbox{CNOT} in this proposed mechanism,  
which can be done by a single controlled rotation by $\pi/2$. 
This can be written as
\begin{equation}
\fbox{CNOT}=e^{-iE_0\,t_2}\, [{\cal C}_2 ~\to ~{\cal C}_1]\,
\,R_x(\pi/2)\, [{\cal C}_2 \to {\cal C}_2] \label{cz.4}~.
\end{equation}
If one ignores the overall phase, Eq.\,(\ref{cz.4}) amounts
to setting the coupling so that ${\cal C}_2={\cal C}_1$, applying
the rf-field to induce the rotation $R_{x}(\pi/2)$,
and finally switching back to the initial
configuration of the ${\cal C}_1$ and ${\cal C}_2$. 
The full \fbox{CNOT} matrix
in the $\vert {\bf v}\rangle$ basis is
\begin{equation}
\fbox{CNOT}=\pmatrix{0 & 0 & 0 & i\cr
                     0 & 1 & 0 & 0\cr
                     0 & 0 & 1 & 0\cr
                     i & 0 & 0 & 0\cr}
\label{cnot.1}
\end{equation}
which amounts to the flip of the second bit $\vert 0\rangle \to
i\vert 1\rangle$ and $\vert 1\rangle \to i\vert 0\rangle$ when the
first bit is in the $1$ state.

\subsection{Operational time scales and dephasing} 
The time scale required for the gate operations can be rougly considered 
to be independent from the particular gate operation. We consider the  
operational time as $t_{op} \simeq (\Delta E)^{-1}$ where $\Delta E$ describes 
the energy difference 
of the states at the configuration where the gate operation is performed. 
For instance, a typical scale is the phase flip 
$t_{op}=t_1=\pi/4{\cal C}_1$ can be 
read off from Fig.\,4 at the degeneracy point $y_{ex}=y_{ex}^*$. We find that 
${\cal C}_1 \sim 5 \, 10^{-3} eV$ which yields $t_1 \sim 5 ps$. The switching 
time is possible for less than $1 ns$ between two gate operations.
\cite{makhlin}
By comparing these two time scales, we estimate that the gate 
operations are dictated by the switching time. 

As the dephasing mechanisms are concerned certain expected sources and 
their decoherence time scales have been previously analyzed in 
detail.\cite{makhlin,mooij} SQUIDs in the flux regime are known to be 
insensitive to the charge fluctuations. The decoherence time for the 
charge fluctuations in the flux dominated regime has been 
estimated\cite{makhlin} to be $t_{\varphi} \sim 0.1 s$. Among the 
other sources there 
are, the quasiparticle tunneling ($t_{\varphi} \sim 1 ms$) and 
electromagnetic 
radiation of the junction\cite{tian} ($t_{\varphi} \sim 10^3 s$). More 
substantial dephasing effects are known to arise from the near distance 
dipole-dipole interactions between the SQUIDs ($t_{\varphi} \sim 0.1 ms$). 

The relaxation time in the time dependence of the basis (flux or parity) 
states is the most crucial element for the I/O operations based on the density 
matrix measurements. For the parameters 
considered here it can be calculated by the explicit expression given by 
Leggett et al. and Weiss Ref.\,[11] for which we find  
$t_{rel} \sim 1 \mu s$.  

\subsection{Read-in and read-out}
The state of the coupled squids is observed before and after the
computation by the flux measurement made by two dc-squids each
independently facing one SQUID loop as shown in Fig.2. Before any
read-in or read-out made, a back transformation is needed between  
the operational $\vert {\bf v}\rangle$ basis to the physical
flux basis $\vert {\bf V} \rangle$. Suppose that a particular
logic operation is denoted by the matrix $T$ in the basis 
$\vert {\bf v}\rangle$. The formula
\begin{equation}
{\bf V} ~\to ~ [{\cal U}_2{\cal U}_1]\,T\,
[{\cal U}_2{\cal U}_1]^{-1}\,{\bf V}
\label{backtrf}
\end{equation}
describes the effect of $T$ in the basis $\vert {\bf V}\rangle$. 
Here $T$ is any desired
composition of the single and double qubit operations.
Suppose $T=\fbox{Z}_{\phi}$. 
Using (\ref{backtrf}) we find that the flux states
are transformed by,
\begin{equation}
\fbox{Z}_{\phi}:\vert {\bf V}\rangle=
\pmatrix{\cos\phi & 0 & 0& i\sin\phi\cr
         0& \cos\phi & i\sin\phi & 0\cr
         0& i\sin\phi & \cos\phi & 0\cr
         i\sin\phi & 0 & 0 & \cos\phi\cr}\,\vert {\bf V}\rangle~.
\label{z.2.2}
\end{equation}
We give a second example from $T=\fbox{CNOT}$. Applying (\ref{backtrf})
to this case we obtain
\begin{equation}
\vert {\bf V}\rangle ~\to ~{1 \over 2}\,
\pmatrix{(1-i) & 0 & 0 & (1+i)\cr
         0 & (1+i) & 0 & -(1-i)\cr
         -(1+i) & 0 & (1-i) & 0\cr
          0 & -(1-i) & 0 & (1+i)\cr}\,\vert {\bf V}\rangle~.
\label{flux.2}
\end{equation}
The unique correspondence between the flux states and the parity
basis permits read-in and read-out operations to be performed by
measuring the flux as shown in Fig.2. The flux measurements
simulate single event quantum mechanics. Suppose an output state
to be measured is $\vert \psi\rangle=a_1\,\vert\uparrow
\uparrow\rangle+ a_2\,\vert\downarrow \uparrow\rangle +
a_3\,\vert\uparrow \downarrow\rangle + a_4\,\vert\downarrow
\downarrow\rangle$
 A single flux measurement yields one of the flux states
in the output with the probability $\vert a_i\vert^2$ associated
with that flux state. The partial amplitudes of the flux states
can therefore be inferred only after repeated measurements. In
this context, the final measurement is similar to the quantum
computation model using ion traps.

\subsection{Discussion}
We demonstrated a new theoretical
mechanism for quantum logic gate operations by
manipulating degeneracy of a symmetric four level system by two variable
coupling parameters. A realization of
this system using rf-SQUIDs in the flux regime is made in which only one
of the couplings is variable. Physical examples in which both 
couplings vary are very likely to exist but currently unknown to us. In that 
case an ideal realization of the proposed mechanism is possible with the 
particularly added feature of Fig.1a namely    
storing the computed information by suppressing the energetic 
interference between the states. The realization of the gates for the 
fundamental single and double qubit operations is also demonstrated. As 
the scalability of this SQUID realization is concerned we have less to 
say at this moment. Masking and evaporation techniques that are already 
well established can be used to fabricate arrays of the qubit pairs. 
The coupling between different pairs can be realized in a similar 
way as those that are coupled as members of a single pair. We continue 
our current effort in the direction of simulating two and three such 
pairs coupled inductively.  

\section*{Acknowledgments}
TH is thankful to I.O Kulik (Department of Physics, Bilkent University) 
for discussions and critical reading of the manuscript.

\section*{figures}

Fig.1: Lifting the four-fold degeneracy stepwise in the proposed
mechanism. In the ordering of the flux states shown in (c)
[see Eq.\,(\ref{updownst.1}) for the definition of the flux states]
with respect to the energy eigenvalues we considered
$0 < {\cal C}_2 \le {\cal C}_1$. Note that the initial four-fold degeneracy
causes the energy centers indicated by the horizontal dashed lines to be
independent of the couplings.

Fig.2: Inductive coupling of the two rf-SQUID loops. The
dc-SQUIDs have two-junctions for balanced
read-in and read-out the flux at each rf-SQUID loop independently.
Although the details of the mutual field screening is not shown,
the circuitry should be designed such that the I/O devices are
decoupled from each other completely. 

Fig.3: Pseudo-degenerate energy eigenfunctions and the corresponding
single well localized flux states. The flux states $0 \pm 1$ are properly
normalized by $1/\sqrt{2}$. The parameters used are $\beta_L=1.1$,
$y_{ex}=\pi$, $C=0.4 pF$, $L=100 pH$.  

Fig.4: Lifting the degeneracy by tilting the symmetric double-well
potential in the coupled SQUID pair. The vertial scale is the
numerically calculated eigen energies in eV units. The horizontal variable
corresponds to no tilt when $y_{ex}=\pi$ and critical tilt when
$y_{ex}=y_{ex}^*$. The tilting parameter $y_{ex}/\pi$ simulates the
coupling ${\cal C}_2$ in the proposed mechanism in section A.
The parameters used are $\beta_L=1.1$, $C=0.4 pF$,
$L=100 pH$ and $M=0.03 L$.

Fig.5: The transformation from the flux basis $\vert {\bf V}\rangle$
to the parity basis by two Hadamard transformations. After ${\cal U}_1$
is applied on the flux basis, the states are
$v_1^\prime=(\vert \uparrow\downarrow\rangle-
\vert\downarrow\uparrow\rangle)/\sqrt{2};
v_2^\prime=(\vert \uparrow\downarrow\rangle+
\vert\downarrow\uparrow\rangle)/\sqrt{2}$; \\
$v_3\prime=(\vert \uparrow\uparrow\rangle-
\vert\downarrow\downarrow\rangle)/\sqrt{2}; ~
v_4^\prime=(\vert \uparrow\uparrow\rangle+
\vert\downarrow\downarrow\rangle)/\sqrt{2}$ with the energies
$E_1=E_0-{\cal C}_1-{\cal C}_2,~E_2=E_0-{\cal C}_1+{\cal C}_2$,~\\
$E_3=E_0+{\cal C}_1-{\cal C}_2,~E_4=E_0+{\cal C}_1+{\cal C}_2$~.

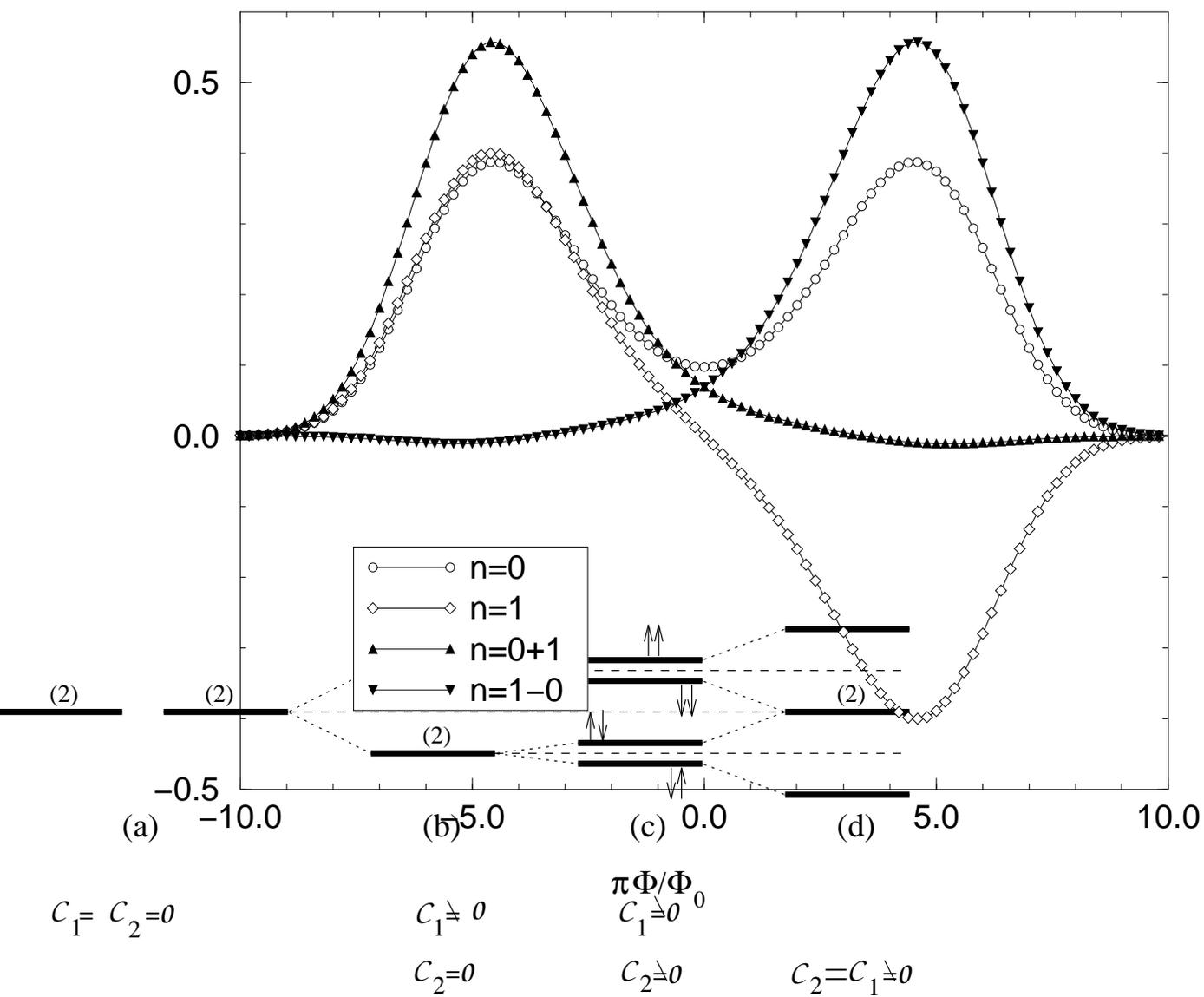
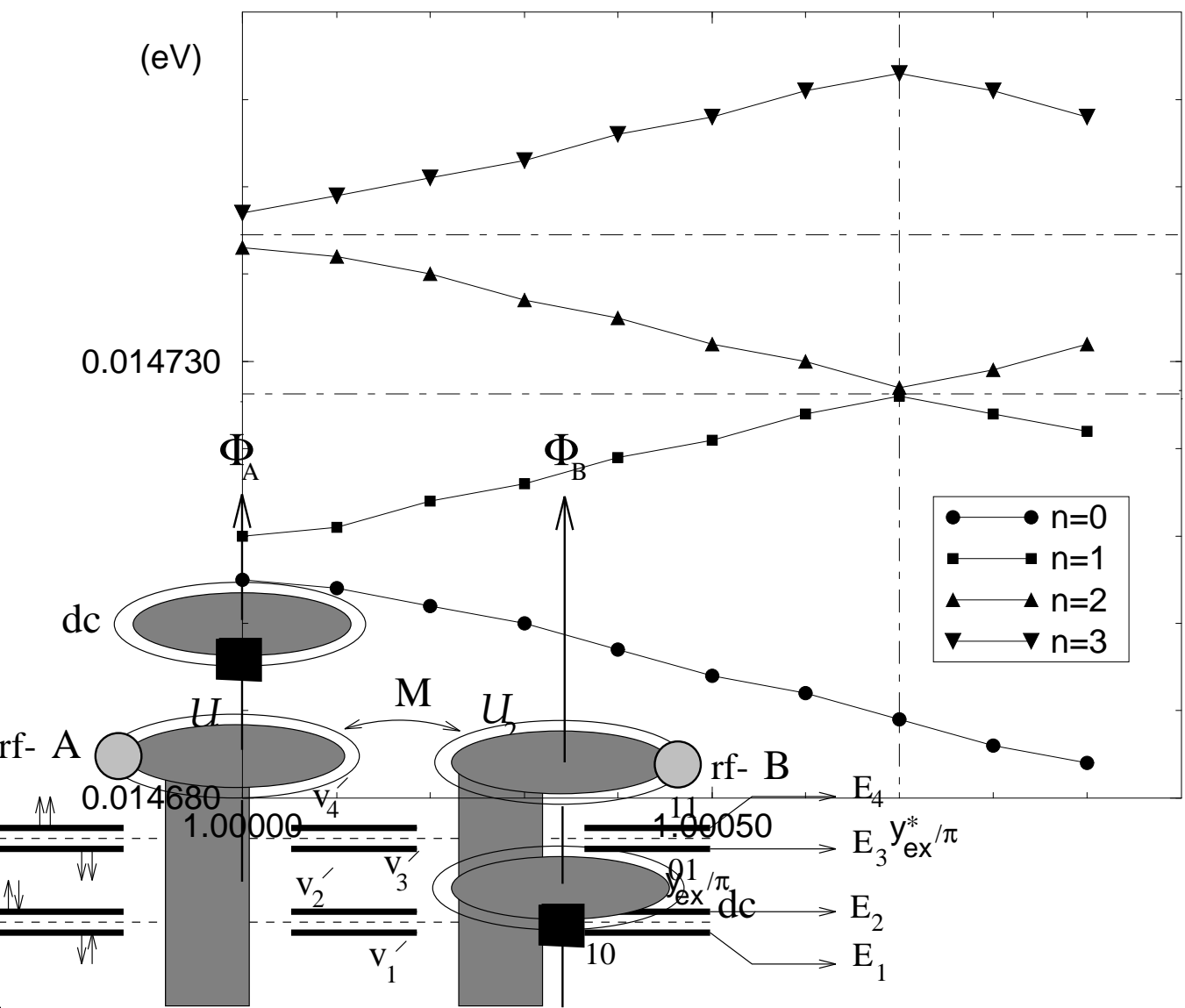
\end{document}